\documentclass[aps,prl,twocolumn,showpacs]{revtex4-1}
\usepackage{graphicx}
\usepackage{bm}
\usepackage{color}

\begin{document}

\title{Coexistence of Anomalous and Normal Diffusion in Integrable Mott Insulators}
\author{R. Steinigeweg$^1$, J. Herbrych$^1$ and P. Prelov\v{s}ek$^{1,2}$}
\affiliation{$^1$J. Stefan Institute, SI-1000 Ljubljana, Slovenia}
\affiliation{$^2$Faculty of Mathematics and Physics, University of Ljubljana, SI-1000 Ljubljana, Slovenia}
\author{M. Mierzejewski$^3$}
\affiliation{$^3$Institute of Physics, University of Silesia, 40-007 Katowice, Poland}

\date{\today}

\begin{abstract}
We study the finite-momentum spin dynamics in the one-dimensional XXZ spin chain within
the Ising-type regime at high temperatures using density autocorrelations within linear response
theory and real-time propagation of nonequilibrium densities. While for the nonintegrable model
results are well consistent with normal diffusion, the finite-size integrable model unveils the
coexistence of anomalous and normal diffusion in different regimes of time. In particular,
numerical results show a Gaussian relaxation at smallest nonzero momenta which we relate to
nonzero stiffness in a grand canonical ensemble. For larger but still small momenta normal-like
diffusion is recovered. Similar results for the model of impenetrable particles also help to resolve
rather conflicting conclusions on transport in integrable Mott insulators. 
\end{abstract}
% 72.10.-d Theory of electronic transport; scattering mechanisms
% 71.27.+a Strongly correlated electron systems; heavy fermions
% 72.10.Bg General formulation of transport theory
% 75.10.Pq Spin chain models
% 75.10.Jm Quantized spin models, including quantum spin frustration
% 05.60.Gg Quantum transport
% 05.70.Ln Nonequilibrium and irreversible thermodynamics

\maketitle

{\it Introduction.}---
Theoretical investigations of transport in many-body systems of interacting fermions established
several novel, entirely quantum aspects which go well beyond usual weak-scattering or
Boltzmann-type approaches to transport. It has been shown that integrability of a model system
can change qualitatively the response to external driving. A prominent experimentally relevant
example of such a system is the one-dimensional (1D) Heisenberg (XXZ) model \cite{hess2007},
where a long-standing question is the existence of diffusion in the isotropic case \cite{thurber2001,
sirker2009, grossjohann2010}, and recently also spin systems mapping on the anisotropic
(Ising-like) case became of interest \cite{kimura2007}.

While it is by now quite well established that integrable conductors, in the easy plane regime,
exhibit at any temperature $T>0$ ballistic (dissipationless) transport within linear response (LR)
theory \cite{zotos1996, review2007} characterized by a finite spin stiffness $D_s(T>0)>0$
\cite{zotos1997, zotos1999, benz2005, herbrych2011}, which was recently confirmed by exact
lower bounds \cite{prosen2011}, $T>0$ transport in the Ising-type (Mott insulating in the
fermionic representation) regime of the same model still represents a challenge with some
apparently conflicting conclusions. While  $D_s(T \ge 0)=0$ for an infinite system in this
regime, numerical studies for the LR dynamical conductivity $\sigma(\omega)$ at high
$T \gg 0$ reveal in finite systems of length $L$ a broad incoherent response with on the one
hand quite featureless spectra in the thermodynamic limit $L \to \infty$ and consequently a finite
d.c.~value $\sigma_\mathrm{dc} =\sigma(\omega \rightarrow 0)$ \cite{prelovsek2004}. On the
other hand, the low-$\omega$ behavior is dominated by a huge finite-size anomaly with vanishing
response within a window $\omega < \omega^* \propto 1/L$. Anomalous behavior reappears also
at finite-field driving or weak perturbation (of the model), which both break the integrability
and indicate on the existence of  an ``ideal insulator'' with $\sigma_\mathrm{dc} \to 0$
\cite{mierzejewski2011} as the proper limit. Also the exact nonequilibrium steady state of a
strongly driven open XXZ chain \cite{prosen2011} reveals a similar anomaly with the current
decaying exponentially with the length of the chain.  This is in contrast with steady transport under 
{\it near--equilibrium} conditions suggesting again a finite diffusion constant ${\cal D}$
\cite{znidaric2011} which is also consistent with previous studies performed at finite time
\cite{steinigeweg2009} and momentum \cite{steinigeweg2011}.

The aim of this Letter is to reconcile apparently inconsistent manifestations of diffusion in the
(Mott) insulating regime at $T>0$, in particular at high $T \to \infty$, whereby we concentrate
our analysis on the XXZ Heisenberg model with the anisotropy $\Delta>1$. We first note that
anomalous transport in the integrable model can be related to the absence of a characteristic scale
representing ``the mean free path'' $l^*$, which is in small systems substituted effectively by the
actual size $L$. On the other hand the diffusion constant ${\cal D}$ in the thermodynamic limit
$L \to\infty$ would indicate a very short $l^* \sim 1$. This dichotomy shows up for finite systems
in spin correlations $S_q(t)$ at momentum $q>0$ as diffusion-type decay at $t<t^*$ while at
$t>t^*$ the decay becomes Gaussian, whereby $t^* \propto L$. The latter behavior is shown to be
dominant at smallest non-zero $q=q_1=2\pi/L$ and its origin can be traced back to the existence 
of a finite stiffness $D_s>0$ in a grand canonical ensemble. With increasing $q > q_1$ normal
diffusion prevails. To strengthen our arguments and results we confirm the same phenomena 
within the 1D model of impenetrable particles ($U \rightarrow \infty$ Hubbard model), where
clearly no steady spin current is possible at zero magnetization, nevertheless $S_{q>0}(t)$
again reveals a coexistence of normal diffusion and Gaussian decay.

{\it 1D Heisenberg model.}---
First, we address the question of spin transport in the 1D anisotropic Heisenberg model,
\begin{equation}
H = J \sum_{r=1}^L (S_r^x S_{r+1}^x + S_r^y S_{r+1}^y + \Delta S_r^z S_{r+1}^z +
\Delta_2 S_r^z S_{r+2}^z) \, , \label{H}
\end{equation}
where $S_r^i$ ($i = x,y,z$) are spin $s=1/2$ operators at site $r$, $L$ the length of the chain
with periodic boundary conditions (p.b.c.), and $\Delta$ represents the anisotropy. We allow also
for a next-nearest neighbor $zz$-interaction with $\Delta_2 \neq 0$ breaking the integrability
of the model. It should be reminded that the Hamiltonian (\ref{H}) can be mapped on a 
$\tilde{t}$--$V$--$W$ model of interacting spinless fermions with hopping $\tilde{t} = J/2$ and
inter-site interactions $V = J \Delta, W = J \Delta_2$. In this fermionic picture, spin transport
corresponds to charge transport and the here interesting regime of  $\Delta>1$,
$S^z_\mathrm{tot}=0$ (at $\Delta_2 \sim 0$) to the Mott insulator.

Our aim is to analyze spin transport in the insulating regime $\Delta>1$ at $T\gg 0$. Two
complementary numerical approaches are used: a) LR theory calculating relevant dynamical
spin correlation functions and b) real-time propagation (TP) of spin after switching off  a
perturbing magnetic field. The latter can be regarded as the control of the validity of LR
theory at finite perturbations, a question being nontrivial in particular for integrable (and
nonergodic) systems.  In both approaches, however, we apply essentially the same numerical
approach by means of Lanczos diagonalization of finite systems \cite{prelovsek2011}, 
applicable up to $L \sim 30$, beyond the range of full diagonalization (FD), where $L \lesssim
20$ \cite{fabricius1998, steinigeweg2011}.

Within the framework of LR theory we consider the time-dependent spin correlation function
$S_q(t) = \mathrm{Re} \langle S_q^ z(t) S_{-q}^ z \rangle/L$ where $S_q^z = \sum_r
e^{\imath q r} S_r^z$ and, due to p.b.c., $q=2\pi k/L$. Here, $\langle \ldots \rangle$ denotes 
the thermodynamic average at temperature $T$.  We primarily focus on the high-temperature
limit $\beta=1/T \rightarrow 0$ at zero magnetization $S^z_\mathrm{tot}=0$ where we use the
microcanonical Lanczos method (MCLM)  \cite{prelovsek2011} to evaluate $S_q(\omega)$ in
finite systems and then perform the Fourier transform into $t$-dependent $S_q(t)$. In the case
of perfectly diffusive dynamics we would expect $S_q(t) \propto \exp(-q^2 {\cal D} t)$. In
general, however, the instantaneous rate ${\cal D}_q(t) = -\dot{S}_q(t)/[q^2 S_q(t)]$
\cite{steinigeweg2011} can become constant only in a hydrodynamic regime at small enough
$q$ and long $t$. For such $q$, ${\cal D}_q(t)$ is related to the autocorrelation function
$J_0(t) = \langle J_0^z(t) J_0^z \rangle/L$ of the $q=0$ current $J_0^z = J \sum_r (S_r^x
S_{r+1}^y - S_r^y S_{r+1}^x)$ by the Einstein relation \cite{steinigeweg2009} (at $\beta
\to 0$)
\begin{equation}
\lim_{q \rightarrow 0} {\cal D}_q(t) = \frac{\sigma(t)}{\chi} = \frac{1}{S_0(t=0)}
\int_0^t \! \mathrm{d}t' \, J_0(t')  \, , \label{einstein}
\end{equation}
assuming non-singular behavior at $q \to 0$. This assumption becomes
relevant since a finite system features a nonvanishing stiffness $D_s$, i.e., $J_0(t > t^*) = 2
D_s$, in particular if one considers the grand canonical averaging (over all $S^z_\mathrm{tot}$
in the XXZ model) where $D_s \propto 1/L$ \cite{review2007} as discussed furtheron.
Hence, due to
\begin{equation}
\lim_{q \rightarrow 0} {\cal D}_q(t > t^*) = \mathrm{const.} + \frac{2 \, D_s \,  t}{S_0(t=0)}
\, , \label{linear}
\end{equation}
a finite stiffness $D_s$ restricts normal diffusion to $t < t^*$ and implies anomalous dynamics
for $t > t^*$ \cite{steinigeweg2009}. 

\begin{figure}[t]
\includegraphics[width=0.85\columnwidth]{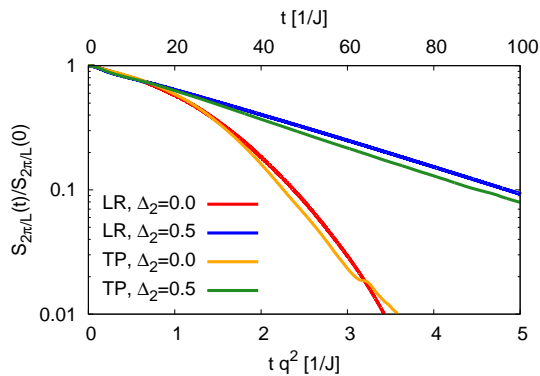}
\caption{(color online)  Decay of spin densities $S_q(t)$ in the anisotropic Heisenberg
model with $\Delta=1.5$ shown for smallest nonzero $q=q_1$ on a chain of length $L=26$ and
$\Delta_2 =0$, $0.5$, corresponding to integrable and nonintegrable models, respectively. LR
results are obtained via the MCLM method at $\beta =0$ while TP results are obtained at
$\beta=0.4/J$. }
\label{Fig1}
\end{figure}

To study spin transport using the real-time dynamics, we extend the Hamiltonian (\ref{H}) by
introducing a position-- and time--dependent magnetic field, $H \rightarrow H- \sum_r  h_r(t)
S^z_r$. The initial equilibrium state corresponding to small but finite inverse temperature $\beta
\ll 1/J$ is then obtained by means of the MCLM \cite{prelovsek2011} with the initial perturbation
$h_r(t \le 0) = h_0 \cos(q r)$. Finite magnetic field induces site--dependent magnetization which
for $\beta \ll 1/J$ is on average $\langle S^z_r \rangle(0) \approx \beta h_r(0)/4$. At $t=0$ the
field is switched off and the system evolution is obtained from the Lanczos time propagation method
\cite{mierzejewski2010, mierzejewski2011, prelovsek2011}. Following the relaxation of local
magnetization $\langle S^z_r \rangle$(t) we first check that the magnetization preserves its initial
spatial profile for the assumed $h_r(t)$, i.e.,  $\langle S^z_r \rangle (t)=\langle S^z_r \rangle(0) f(t)$
even when the dynamics are anomalous and strongly deviate from a simple exponential
diffusion dependence $\langle S^z_r \rangle(t)=\langle S^z_r \rangle(0) \exp(-{\cal D}q^2t)$.
Therefore, $f(t)$ can be used for distinguishing between normal diffusion and anomalous dynamics.
For $h_0 \rightarrow 0$ this quantity should be compared with the ratio $S_q(t)/S_q(0)$ from the
LR approach. Such a comparison for smallest finite $q=q_1$ is shown in Fig.~\ref{Fig1}.  A
quantitative agreement between both methods is clearly visible, confirming that the LR approach
(for $q>0$) remains for finite $h_0$ valid even when the system is integrable. On the other hand,
the key result in Fig.~\ref{Fig1} concerns clear presence of normal diffusion in a generic
nonintegrable system ($\Delta_2 \ne 0$) in sharp contrast with fast and anomalous relaxation
visible in the integrable system.
 
% ------------------------------- FIGURE1 ------------------------------------
%-----------------------------------------------------------------------------

% ------------------------------- FIGURE2 ------------------------------------
\begin{figure}[b]
\includegraphics[width=0.85\columnwidth]{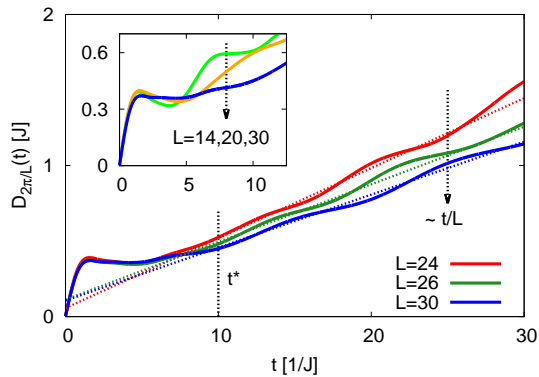}
\caption{(color online) Spin decay rate ${\cal D}_q(t)$ at anisotropy $\Delta=2.0, \Delta_2=0$
calculated for the smallest finite $q_1$ in chains of different $L=14, 20, 24, 26, 30$. Dashed
curves indicate straight lines $\propto 1.0 \, t J^2/L$ while $t^*=L/(3J)$ is marked for $L=30$.
The inset emphasizes the $L$--dependence at short times.}
\label{Fig2}
\end{figure}
%-----------------------------------------------------------------------------

Next we focus on the behavior of the integrable case with $\Delta_2=0$. To give insight into the 
origin of the fast decay of $S_q(t)$ we consider the instantaneous rate ${\cal D}_q(t)$. In
Fig.~\ref{Fig2} we thus summarize our numerical LR results on ${\cal D}_q(t)$ at $q = q_1$
and $\Delta = 2.0$ for chains ranging from $L=14$ to $30$, where FD is used for $L \leq 20$
and the MCLM for $L=24 $--$30$. Apparently, ${\cal D}_q(t)$ first increases at short times
$t \lesssim 1.5/J$ and then develops a rather constant plateau ${\cal D}_q(t) \approx 0.4 J$ at
intermediate times $t < t^*$, which is consistent with previous studies at finite momentum yielding
${\cal D}_q(t) \approx 0.88 J/\Delta$ \cite{steinigeweg2011}. Clearly, the plateau marks diffusive
dynamics at intermediate time scales and we further observe this time scale to increase with system
size approximately as $t^* \approx L/(3J)$, see Fig.~\ref{Fig2} (inset). While this scaling with
$L$ is a pointer to purely diffusive dynamics in the thermodynamic limit $L \rightarrow \infty$,
the dynamics at long times $t > t^*$ turns out to be different for finite $L$, in particular for
$q=q_1$. As clearly visible in Fig.~\ref{Fig2}, ${\cal D}_q(t)$ increases linearly with time, which
indicates anomalous dynamics. One might be tempted to relate the linear increase directly to a finite
stiffness $D_s$, cf.~Eq.~(\ref{linear}).  But in a canonical ensemble at zero magnetization the
stiffness decreases exponentially fast with system size \cite{prelovsek2004} while the slope in
Fig.~\ref{Fig2} scales rather as $1/L$. Hence, we compare the slope with the stiffness resulting for
a grand canonical ensemble (over all $S^z_\mathrm{tot}$ sectors) with zero average magnetization.
Since lower bounds for $D_s$ are given by the Mazur inequality, and a projection to the conserved
energy current has been shown to represent well the actual stiffness \cite{herbrych2011}, we arrive at
\begin{equation}
D_s^\mathrm{gc} \geq \frac{\Delta^2 \, J^2}{4 \, (1 + 2 \Delta^2) \, L},
\end{equation}
and, noting the sum rule $S_0(t=0) = 1/4$, we obtain $2 D_s^\mathrm{gc}/S_0(t=0) \geq 0.89 J^2/L$
for $\Delta=2.0$ while FD results yield $\approx 1.19 J^2/L$ for small system sizes \cite{review2007,
steinigeweg2009}. The convincing agreement with the slope of ${\cal D}_q(t)$ in Fig.~\ref{Fig2},
$\approx 1.0 J^2/L$, is a hint at an effective finite stiffness at nonzero $q=q_1$, not being reported yet.
In any case, the linear increase of the rate ${\cal D}_q(t)$ at longer times $t > t^*$ identifies the fast
Gaussian decay of $S_q(t)$ in Fig.~\ref{Fig1}, similarly as found for transport in complex one-particle
models of finite size \cite{steinigeweg2007}. Even though not shown here explicitly, this type of relaxation
also manifests in the spectrum $S_q(\omega)$ as an anomaly of Gaussian shape at low frequencies $\omega
< \omega^* = 2\pi/t^* \propto 1/L$, well pronounced at $q=q_1$ since in this case the main part of the
sum rule is located at $\omega<\omega^*$.

% ------------------------------- FIGURE3 ------------------------------------
\begin{figure}[t]
\includegraphics[width=0.85\columnwidth]{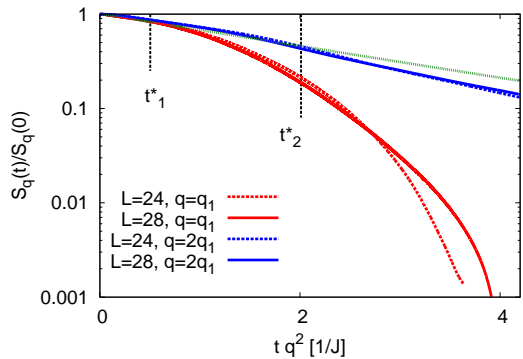}
\caption{(color online) Decay of densities $S_q$ vs.~scaled time $t q^2$ at anisotropy
$\Delta=2.0$, calculated within LR via MCLM on chains with $L= 24, 28$ sites for different
wave vectors $q = q_1, 2q_1$, respectively. Dashed curve indicates exponential decay with
${\cal D}=0.4 J$. The position of the marginal time $t^*$ ($=t^*_{1,2}$) is also marked for
$L=28$.}
\label{Fig3}
\end{figure}
%-----------------------------------------------------------------------------

So far, we have concentrated on $S_q(t)$ at the smallest finite $q=q_1$. Here, the effective
relaxation time $\tau_q \gg t^*$ and the exponential decay only appears as a minor fraction
of the total relaxation while Gaussian relaxation dominates, see Fig.~\ref{Fig3}. On the other hand, 
for large enough $q>q_1$, we realize that $\tau \lesssim t^*$ and the exponential relaxation starts
to dominate the decay curve. Thus, a pertinent criterion for ``normal'' diffusion relaxation is given
by $\tau = 1/(q^2 {\cal D}) \lesssim t^*$. The latter criterion already turns out to be rather well satisfied
for $q =2 q_1$ for the considered chain lengths, e.g., for $L = 28$, as illustrated in Fig.~\ref{Fig3}. 

{\it 1D model of impenetrable particles.}---
We are finally going to address the question to which extent the observed dynamical behavior are
generic for integrable quantum systems in the insulating (Mott-Hubbard-type) regime. To this end
we investigate the 1D model of impenetrable particles, which has also been shown to behave
anomalously with respect to transport \cite{prelovsek2004}. The model (being the 
$U \rightarrow \infty$ limit of the Hubbard model or $J \to 0$ limit of the $\tilde{t}$--$J$ model)
is given by the Hamiltonian

\begin{equation}
H = -\tilde{t} \sum_{r=1}^L \sum_{s} \tilde{c}_{r+1,s}^\dagger \tilde{c}_{r,s} + \mathrm{H.c.}
\label{H2},
\end{equation}
where projected fermion operators $\tilde{c}_{r,s}=c_{r,s}(1-n_{r,-s})$ take into account that
double occupancy of sites is forbidden. The two different species of particles are given by up
($\uparrow$) and down ($\downarrow)$ spin fermions. We should note that there is a close
analogy of the XXZ model (\ref{H}) in the large anisotropy (Ising) limit $\Delta \gg 1$ with the
$\tilde t$ model (\ref{H2}). Namely, within the Ising limit we are dealing with the Ne\'{e}l
ordered ground state (at $S^z_\mathrm{tot}=0$) and the excited states composed of split
subspaces of oppositely charged ``soliton-antisoliton" ($s \bar s$) pairs. In such a limit, the
solitons/antisolitons behave effectively as impenetrable quantum particles since their crossing
would require virtual processes with energy $\delta E= J \Delta$ within the XXZ model (or
$\delta E =U$ within the Hubbard model).

\begin{figure}[t]
\includegraphics[width=0.85\columnwidth]{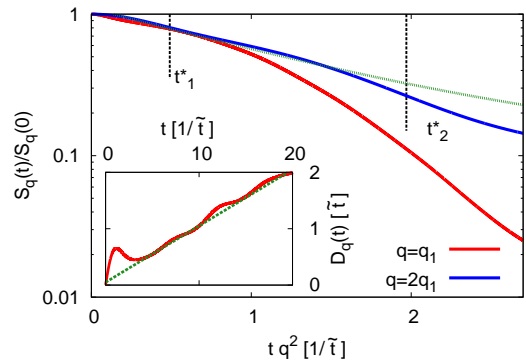}
\caption{(color online) Decay of spin density autocorrelations $S_q(t)$ in the $\tilde{t}$ model at
$n=1/2$ (quarter filling), zero magnetization, and high temperatures, depicted for the two smallest
momenta $q>0$ in a chain of length $L=20$. One solid curve indicates an exponential
decay with a constant rate ${\cal D} = 0.76 \tilde{t} $. Inset: Rate ${\cal D}_q(t)$ at the smaller of
both $q$ for the same parameter set. The dashed curve indicates a line $\propto 0.1 \, t \,
\tilde{t}^2$.}
\label{Fig4}
\end{figure}

Within the $\tilde t$ model, the charge and spin currents can be written as
\begin{equation}
J_0^{[c, s]}= \tilde{t} \sum_{r=1}^L \sum_{s} \imath [1, s] \, \tilde{c}^\dagger_{r+1,s}
\tilde{c}_{r,s}+  \mathrm{H.c.}
\end{equation}
with $\langle J_0^{[c,s]} J_0^{[c,s]} \rangle/L = [2,1/2] \, n (1-n) \tilde{t}^2$, where 
$n = (N_\uparrow + N_\downarrow)/L$ is the filling. Further we notice that $J_0^c$ commutes with $H$ 
(from the
perspective of charge the model is equivalent to 1D noninteracting spinless fermions) while
$J_0^s$ does not. Their overlap $\langle J_0^c J_0^s \rangle/L = 2 m (1-n) \tilde{t}^2$ vanishes
when magnetization $m = (N_\uparrow - N_\downarrow)/(2 L)$ is zero. Moreover, it is quite
evident that for $m=0$ there could be no d.c.~spin transport since $N_\uparrow$ particles cannot
cross with $N_\downarrow$ particles which implies $D_s=0$ but as well
$\sigma_\mathrm{dc} = \sigma(\omega \to 0)=0$ \cite{prelovsek2004}. On the other
hand, using a grand-canonical ensemble with average $m=0$, the Mazur inequality leads to the
lower bound
\begin{equation}
D_s^\mathrm{gc} \geq \frac{(1-n)  \, \tilde{t}^2}{4 \, L} \, .
\end{equation}
Noting the sum rule $S_0(t=0) = n/4$, we obtain $2 D_s^\mathrm{gc}/S_0(t=0) \geq 2.0 \tilde{t}^2/L$
for $n=1/2$ (corresponding to quarter filling for the Hubbard model), as considered in the following.
This lower bound we again compare with the instantaneous rate ${\cal D}_q(t)$ at smallest $q=q_1$
in a chain of length $L=20$, maximally treatable with MCLM. As shown in Fig.~\ref{Fig4} (inset),
we indeed find ${\cal D}_q(t)$ to increase remarkably well linearly at long times $t > t^*$, as before,
in obvious agreement with the lower bound. This linear increase leads to a Gaussian decay of $S_q(t)$
at long times $t > t^*$. On the other hand, frequency moments and the limit $L \rightarrow \infty$ of
the dynamical spin conductivity $\sigma(\omega)$ are again consistent with a finite diffusion constant
being in this ($n=1/2$) case ${\cal D}= 0.76 \tilde{t}$ \cite{prelovsek2004}. Also we observe the
dynamics at $t < t^*$ to be consistent with an exponential decay involving this value for the diffusion
coefficient, as shown in Fig.~\ref{Fig4}. As well we confirm in Fig.~\ref{Fig4} that for larger $q=2q_1$
the decay approaches the ``normal'' diffusion behavior.

% ------------------------------- FIGURE4 ------------------------------------
%-----------------------------------------------------------------------------

{\it Conclusion.}---
In summary we studied the finite-$q$ spin dynamics in the 1D Heisenberg chain with anisotropy
$\Delta > 1$ in the high-temperature limit $\beta \to 0$. As one of the main results, we first showed the
validity of linear response theory at finite perturbations using the real-time propagation of nonequilibrium
densities. While we found exponential relaxation (normal diffusion) in the nonintegrable model, we
observed in the integrable model the coexistence of a Gaussian relaxation (anomalous diffusion) at long
times $t > t^* \propto L$, being dominant  at smallest  $q =q_1$ where the effective relaxation time of spin
modulations is $\tau_q \gg t^*$. On the other hand, when increasing $q > q_1$, normal diffusion prevailed
and also the respective diffusion constant is in quantitative agreement with transport coefficients from steady
state scenarios \cite{znidaric2011}. To be in full agreement with the latter (open system) scenarios it is
therefore important to perform the limits in the appropriate order \cite{buragohain1999}, i.e., first
$L \to \infty$ ($t^* \to \infty$) and then $q \to 0$ ($\tau_q \to \infty)$, although the opposite limits can be
as well relevant and realized \cite{mierzejewski2011}. Finally, we obtained similar results on the 1D model
of impenetrable particles, suggesting that the observed dynamics is quite generic for integrable Mott insulators.

\acknowledgements

This work has been supported by the Program P1-0044 of the Slovenian Research Agency
(ARRS) and RTN-LOTHERM project. M.M.~acknowledges support from the N N202052940
project of MNiSW.

\end{document}